# A Method for Near-Equilibrium Discrete-Velocity Gas Flows

B. T. Nadiga & D. I. Pullin
Graduate Aeronautical Laboratories
California Institute of Technology
Pasadena CA 91125, USA

**Abstract**

We present a simulation scheme for discrete-velocity gases based on *local thermodynamic equilibrium*. Exploiting the kinetic nature of discrete-velocity gases, in that context, results in a natural splitting of fluxes, and the resultant scheme strongly resembles the original processes. The kinetic nature of the scheme and the modeling of the *infinite collision rate* limit, result in a small value of the coefficient of (numerical)-viscosity, the behavior of which is remarkably physical [18]. A first order method, and two second order methods using the total variation diminishing principle are developed and an example application presented. Given the same computer resources, it is expected that with this approach, much higher Reynold's number will be achievable than presently possible with either lattice gas automata or lattice Boltzmann approaches. The ideas being general, the scheme is applicable to any discrete-velocity model, and to lattice gases as well.

1.  **Introduction**

A discrete-velocity model of a fluid was considered as early as 1890 by Maxwell [1] in the context of kinetic theory of gases. Broadwell, in 1964, [2] used it for the first time in a flow situation, in the context of a Couette flow (and soon after to examine the shock structure [3]). While these models were used in other flow calculations subsequently [4,5,6,7,*etc.*], the large scale utilization of these models as a simulation strategy for fluids had to await the work of Wolfram in 1986 [8], in which the computer-scientific idea of a cellular automaton (CA) as described by von Neumann [9], was unified with the idea of a discrete-velocity gas to give rise to what has been called a lattice gas. Since, as is well known, lattice gases have become a field of study in themselves, see [10,11] and references therein. A lattice gas involves, in addition to a discretization of the velocity space, a discretization of the physical space as well, rendering the phase space fully discrete. Such a complete discretization of the phase space of a lattice gas allows it to be implemented on a digital computer as a cellular automaton; such implementations are termed lattice gas automata (LGA). Various shortcomings of LGA have been identified, important of which are those associated with the effective stochastic nature of the scheme resulting in a high level of noise in the simulations [12], and those associated with the conservation of staggered momenta [13,14]. To alleviate these problems, schemes based on the partial-differential equations approximating the lattice gas behavior have been proposed [13,15]. Since the partial-differential equations involved are the particle density conservation equations, much like the classical Boltzmann equations, these methods are termed lattice Boltzmann equation (LBE) techniques. The exact collision terms in the LBEs make solving the LBEs very computation intensive [26], leading to a use of model collision terms which are easier to solve. A popular collision model [15,30] is the single-time relaxation model, in which the physical process of collision is replaced by the tendency of the particle-velocity distribution function towards a known equilibrium, the tendency being proportional to the deviation from the equilibrium. Studies of fluid phenomena using both the above lattice gas techniques (LGA and LBE techniques) abound, see [10,11] and references therein. While the LBE techniques resolve to a certain degree the problems associated with LGA, certain other new issues arise in their usage, *e.g.*, the sufficiency of the single-time relaxation model for molecular collisions. With respect to the LBE techniques, we comment that

(1) One tailors these models to satisfy the incompressible Navier-Stokes equations in principle, while it is known that the models are inherently compressible. (2) In going to the partial differential equation approximation of a lattice gas, space and time are continuous; the *exclusion principle* of LGA — At any time, there can be no more than one particle of a given velocity at a given lattice site — is unnecessary. (Its usage in LGA is primarily due to its computational advantages.) The LBE techniques, however, use the Fermi-Dirac velocity distribution function resulting from such an *exclusion principle* [15]. The second of these objections is easily resolved by replacing the Fermi-Dirac distribution by the corresponding Boltzmann distribution. All of this, however, points to the necessity of a good physics-based simulation technique for discrete-velocity gases — one which retains the basic collisional processes, if only to understand the physics of these models better. We go on to develop one such technique in this paper.

Simplifications aside, discrete-velocity gases are Lagrangian schemes for simulating fluid phenomena, much like molecular dynamics. However, calculation of exact molecular interactions is computationally intensive, perhaps unnecessary, and limits the maximum achievable Reynold's number. It is felt that a better simulation technique for discrete-velocity gases can be achieved by using in conjunction with the underlying concept of particles, an Eulerian picture of the flow field. The resulting scenario would be one in which *cells* in a flow field, constituted by a discrete-velocity gas, interact in terms of mass, momentum, and energy fluxes at the boundaries, with the currency of interactions being the particles with the discrete-velocities. Further, if the interactions between the cells are based on the *local equilibrium* fluxes, then we have a method which is kinetic based and at the same time models the *infinite collision rate* (near-equilibrium) limit of the kinetic process. These ideas come from the equilibrium flux method (EFM) of Pullin [16], which is a kinetic-theory-based finite volume method for calculating the flow of a compressible ideal gas. In that method, the particle ensemble simulation of an ideal gas is replaced by a scheme in which adjacent cells in a cell network interact directly through exchange of mass, momentum, and energy, the exchange in $\Delta t$ being calculated using the equilibrium flux relations [16]. The kinetic flux-splitting aspects of EFM were discussed by Deshpande [17].

In Sec. 2, discrete-velocity gases and lattice gases are described briefly, and expressions developed for the equilibrium fluxes. Sec. 3 discusses the main ideas and develops the technique in the context of a 1-D flow. While Sec. 3.1 presents the algorithm for a first order scheme and interprets it physically, Sec. 3.2 analyses the numerical viscosity of that scheme. Using ideas of Total Variation Dimunition [19,20], Sec. 3.3 develops two second order schemes, a *flux-limited* scheme and a *slope-limited* scheme. The development of the technique is rounded off with a discussion in Sec. 3.4 of issues associated with integration in time. In Sec. 5, numerical implementations of the first order scheme and the two second order schemes, are used on a test problem to establish the spatial accuracies of the three schemes, and Sec. 6 presents the computation of the shock tube problem with the flux-limited second order scheme. While the developments in Secs. 2 & 3 are general and therefore applicable to any discrete-velocity gas, Secs. 5 & 6 use the nine-velocity gas [21,22,23] briefly described in Sec. 4, for the specific computations. Finally, in Sec. 7, the relationship of the present techniqes to LGA and LBE is discussed.

## 2. Equilibrium Fluxes

In its simplest form, a discrete-velocity gas is an ensemble of identical hard sphere particles, with the particles taking on one of a small set of velocities which are defined *a priori*. The evolution of such a gas is exactly like that of a perfect gas and involves free-flight and collision of particles. As mentioned previously, in a lattice gas the physical space is discretized too, and particles are allowed to reside only at lattice sites. This gives rise to a notion of discretized time: particles hop from lattice site to lattice site, commensurate with their velocities, and collide with other particles instantaneously at lattice sites. (The unit of time is that taken by the slowest moving particle to span a lattice link in the direction on its velocity.) The reader is referred to the extensive literature on discrete-velocity gases and lattice gases [4,10] for a detailed description of these models.

Thermodynamic equilibrium in a discrete-velocity gas is defined as the state in which there is a detailed balancing of collisions. It can be shown [18] that, if there are $b$ allowed discrete velocities $(\mathbf{c}_1, \mathbf{c}_2, \cdots, \mathbf{c}_b)$, in general comprising more than one speed, in $D$ spatial dimensions, a thermodynamic equilibrium specified by $D+2$ hydrodynamic parameters, the mass, momentum, and energy, would give rise to a set of $b - D - 2$ relations in that state. For regular discretizations of the velocity space, these relations are of the form

$$n_a n_b \cdots = n_p n_q \cdots \tag{1}$$

where $n_a$ is the probability of a particle having a velocity $\mathbf{c}_a$, *etc*. The exact form of the equations, as to, *e.g.*, whether they are quadratic or cubic, is dependent on the model [18]. While the techniques in this paper are developed in the context of discrete-velocity gases, they are valid for a lattice gas under a change of variables in the equilibrium equations above: the equilibrium equations to be used for a lattice gas are

$$\hat{n}_a \hat{n}_b \cdots = \hat{n}_p \hat{n}_q \cdots, \quad \text{where} \quad \hat{n}_a = \frac{n_a}{1 - n_a}. \tag{2}$$

This change of variables accounts for the effects of *exclusion*, but does not reflect the effects of conservation of staggered momenta.

The vector $\mathbf{F}$ of hydrodynamic variables comprising mass, momentum, and energy is given by

$$\mathbf{F} = \left( \sum_{a=1}^{b} n_a, \sum_a n_a \mathbf{c}_a, \sum_a n_a \mathbf{c}_a^2 \right) \tag{3}$$

and $\mathbf{G}$, the flux of $\mathbf{F}$ is given by

$$\mathbf{G} = \left( \sum_a n_a \mathbf{c}_a, \sum_a n_a \mathbf{c}_a \mathbf{c}_a, \sum_a n_a \mathbf{c}_a^2 \mathbf{c}_a \right) \tag{4}$$

The definition of $\mathbf{F}$ along with the set of $b - D - 2$ thermodynamic equilibrium equations (1) (model dependent), is the *implicit* discrete Maxwell-Boltzmann distribution. Similarly, the definition of $\mathbf{G}$ used in conjunction with the equilibrium equations (1) gives the equilibrium fluxes of mass, momentum, and energy. Since the $b$ particle populations satisfy the $b - D - 2$ equilibrium relations (1), there are $D + 2$ independent particle populations; we represent these $D + 2$ independent particle populations by $\mathbf{m}$. (3) and (4) may then be rewritten as

$$\mathbf{F} = \mathbf{F}(\mathbf{m}) \quad \& \quad \mathbf{G} = \mathbf{G}(\mathbf{m}). \tag{5}$$

While from (3) and (4), it is clear that $\mathbf{G} = \mathbf{G}(\mathbf{F})$, the functional dependence cannot be expressed explicitly. We note that using the implicit forms of $\mathbf{F}$ and $\mathbf{G}$ allows us to work with the *exact* form of the equation of state without having to define it explicitly. Finally, if the equilibrium distribution of all the $b$ populations is denoted by $\mathbf{n}$, $\mathbf{n}$ can be calculated from $\mathbf{m}$, using (1).

We comment, for purposes of shock jump relations which arise in the context of the example application, that the model Euler equations can be written in terms of $\mathbf{F}$ and $\mathbf{G}$ as

$$\frac{\partial \mathbf{F}(\mathbf{m})}{\partial t} + \frac{\partial \mathbf{G}(\mathbf{m})}{\partial x} = 0. \tag{6}$$

## 3. Equilibrium Flow in 1-D

The near-equilibrium flow technique is best illustrated by considering it in one spatial dimension. Consider a linear array of cells tiling the one-dimensional domain. Each cell has a centroid and is bounded by two boundary elements, across which the cell interacts with its neighbors. The time evolution of the system is then reduced to a calculation at each time step of the net flux of $\mathbf{F}$ at the cell boundaries, and updating $\mathbf{F}$ using the fact that $\mathbf{F}$ is a conserved quantity. It is enough to consider the interactions of one cell at one of its boundaries because the domain is invariant under a translation by the dimension of a cell. Considering the interactions at the cell boundary at $x + \Delta x/2$, between the cells centered at $x$ and $x + \Delta x$, the flux of $\mathbf{F}$ at $x + \Delta x/2$, $\mathbf{G}(x + \Delta x/2)$, comes from

1. the flux of $\mathbf{F}$ in the positive $x$-direction due to particles moving in the positive $x$-direction and presently in the cell $[x - \Delta x/2, x + \Delta x/2]$, called $\mathbf{G}^+(\mathbf{n}^+(x + \Delta x/2, t))$ where $\mathbf{n}^+$ has been used to denote the distribution of particles with a positive $x$-velocity, and

2. the flux of $\mathbf{F}$ in the negative $x$-direction due to particles moving in the negative $x$-direction and presently in the cell $[x + \Delta x/2, x + 3\Delta x/2]$, called $\mathbf{G}^-(\mathbf{n}^-(x + \Delta x/2, t))$.

The integral form of the conservation law for $\mathbf{F}$ over the cell $[x - \Delta x/2, x + \Delta x/2]$ centered at $x$ can then be written as

$$\frac{d}{dt} \int_{x-\frac{\Delta x}{2}}^{x+\frac{\Delta x}{2}} \mathbf{F}(\mathbf{n}(x,t))dx + \left( \mathbf{G}^+(\mathbf{n}^+(x + \frac{\Delta x}{2}, t)) - \mathbf{G}^-(\mathbf{n}^-(x + \frac{\Delta x}{2}, t)) \right) \\ - \left( \mathbf{G}^+(\mathbf{n}^+(x - \frac{\Delta x}{2}, t)) - \mathbf{G}^-(\mathbf{n}^-(x - \frac{\Delta x}{2}, t)) \right) = 0. \tag{7}$$

This is the master equation, so to speak, of the present near-equilibrium method and schemes of different orders of accuracy are derived as approximations of this equation. Note that this equation embodies the important physical idea of kinetic flux-splitting, *i.e.*, it is *important* to interpret the flux terms in (7) the way they were introduced earlier. To further congeal the important aspects of this near-equilibrium method, we first discuss the basic first order scheme. The complications arising out of the higher order accurate schemes are discussed thereafter. In that context, the behavior of the first order scheme is seen to be crucial; so while almost always higher order methods are used in computations, an understanding of the first order method is of great importance.

- 6-## 3.1 A First Order Scheme

To obtain a first order scheme, it is sufficient to assume that the velocity distribution, and consequently all other relevant quantities, are constant in the volume of a cell, and that they undergo discontinuous changes at the cell boundaries. The *smallest* length over which any of the quantities in such a setup change is the cell-size, and therefore, one may equivalently say that the mean-free-path of the gas is of the order of the cell size. Considering cell $[x - \Delta x/2, x + \Delta x/2]$ with exchange of mass, momentum, and energy at $x - \Delta x/2$ and $x + \Delta x/2$, the resulting updating scheme for $\mathbf{F}$ in the cell is given by

$$\frac{d}{dt}\mathbf{F}(\mathbf{n}(x,t)) = -\frac{1}{\Delta x}\left\{\mathbf{G}^+(\mathbf{n}^+(x,t)) - \mathbf{G}^-(\mathbf{n}^-(x+\Delta x,t)) \right. \\ \left. - \mathbf{G}^+(\mathbf{n}^+(x-\Delta x,t)) + \mathbf{G}^-(\mathbf{n}^-(x,t))\right\}, \quad (8)$$

where the particles with a positive $u$-velocity at $x + \Delta x/2$ are assumed to be due to the cell $[x - \Delta x/2, x + \Delta x/2]$, and so on. Using a first order time integrator, the forward Euler stepper, the above equation becomes

$$\mathbf{F}(\mathbf{n}(x,t+\Delta t)) = \mathbf{F}(\mathbf{n}(x,t)) \\ - \frac{\Delta t}{\Delta x}\left\{\mathbf{G}^+(\mathbf{n}^+(x,t)) - \mathbf{G}^-(\mathbf{n}^-(x+\Delta x,t)) - \mathbf{G}^+(\mathbf{n}^+(x-\Delta x,t)) + \mathbf{G}^-(\mathbf{n}^-(x,t))\right\}, \quad (9)$$

with $\Delta t$ satisfying the Courant-Fredrich-Levy (CFL) stability criterion $v\Delta t/\Delta x \leq 1$, $v$ being the speed of the fastest particle in the direction considered. The scheme indicated in (9) has a simple physical interpretation in terms of the interactions of the centroids: The state of $x$ at time $t + \Delta t$ is different from the state of $x$ at time $t$ by

1. the departure of particles with a non-zero $u$-velocity from $x$, terms 1 and 4 in (9).

2. the arrival of particles with a positive $u$-velocity from $x - \Delta x$, term 3 in (9) and

3. the arrival of particles with a negative $u$-velocity from $x + \Delta x$, term 2 in (9).

The important difference from the usual discrete-velocity or lattice gas evolution however, is the fact that *the arrival and departure of the particles is so as to simulate fluxes with purely equilibrium components* (no viscous or heat conducting components). Finally, since the primary dependent variables are $\mathbf{m}$, the evolution of $\mathbf{m}$ is given by

$$\mathbf{m}(x,t+\Delta t) = \mathbf{m}(x,t) \\ - \frac{\Delta t}{\Delta x}[\mathbf{J_{Fm}}]^{-1}\left\{\mathbf{G}^+(\mathbf{n}^+(x,t)) - \mathbf{G}^-(\mathbf{n}^-(x+\Delta x,t)) - \mathbf{G}^+(\mathbf{n}^+(x-\Delta x,t)) + \mathbf{G}^-(\mathbf{n}^-(x,t))\right\} \quad (10)$$

where $\mathbf{J_{Fm}}$ is the Jacobian of the transformation from $\mathbf{F}$ to $\mathbf{m}$. Note that the dimension of $\mathbf{m}$, the vector of particle populations that is updated at each time step, here is $D + 2$ and not $b$, as for the usual full discrete-velocity or lattice gas evolution.

## 3.2 Numerical Viscosity of the First Order Scheme

At the outset, it should be mentioned that the scheme outlined in the previous subsection — the cells, fluxes, and everything else — is interpretable as a purely kinetic process, independent of its usage as an approximation. In such an interpretation, the resultant viscosity is no more a mere numerical artifact, but is a direct result of the kinetic process. In part, it is this aspect of the numerical viscosity which contributes to the success of the scheme. The usage of the term *viscosity* in this paper is in the sense of its macroscopic aspects rather than how it comes about. The same holds for mean-free-path. Since the coefficient of viscosity is specific to each model, a qualitative procedure for estimating this coefficient of viscosity is outlined here. Though the particle velocity distributions in the cells themselves are the equilibrium discrete Maxwell-Boltzmann distributions, the distributions at the cell boundaries are not — they are a combination of the two different one-sided equilibrium distributions. At the interface between cell $j$ and $j+1$, indicated by $j+1/2$,

$$\mathbf{n}_a(j + \frac{1}{2}) = \mathbf{n}^+(j) \bigcup \mathbf{n}^-(j+1) \bigcup \mathbf{n}_a^0(j + \frac{1}{2}). \tag{11}$$

The subscript $a$ stands for actual, as opposed to equilibrium which is what is implied by no subscript. The $\mathbf{n}_a$ at $j + 1/2$ defines a macroscopic state there, $\mathbf{F}(j + \frac{1}{2})$. Corresponding to that macrostate, there exists an equilibrium distribution $\mathbf{n}(j + \frac{1}{2})$ and consequently an equilibrium flux $\mathbf{G}(j + \frac{1}{2})$, which can be written as

$$\mathbf{G}(j + \frac{1}{2}) = \mathbf{G}^+(\mathbf{n}^+(\mathbf{F}(j + \frac{1}{2}))) - \mathbf{G}^-(\mathbf{n}^-(\mathbf{F}(j + \frac{1}{2}))) \tag{12}$$

This is, however, not the actual flux at $j + 1/2$. The actual flux at $j + 1/2$, $\mathbf{G}_a(j + \frac{1}{2})$ is given by

$$\mathbf{G}_a(j + \frac{1}{2}) = \mathbf{G}_a(\mathbf{n}_a(j + \frac{1}{2})) = \mathbf{G}^+(\mathbf{n}^+(j)) - \mathbf{G}^-(\mathbf{n}^-(j+1)) \tag{13}$$

where the actual (non-equilibrium) flux at $j + 1/2$ has been expressed in terms of the equilibrium (one-sided) fluxes at $j$ and $j+1$. The non-equilibrium part of the actual flux at $j+1/2$ is given by the difference of (13) and (12). After some manipulation, the non-equilibrium component of the actual flux at $j+1/2$, denoted by $\mathbf{G}_v(j + \frac{1}{2})$, is given correct to first order by

$$\mathbf{G}_v(j + \frac{1}{2}) = -\frac{\Delta x}{2} \left[ (\mathbf{J}_{\mathbf{G}^+\mathbf{m}} + \mathbf{J}_{\mathbf{G}^-\mathbf{m}}) [\mathbf{J}_{\mathbf{Fm}}]^{-1} \frac{d\mathbf{F}}{dx} \right]_{j+\frac{1}{2}} \tag{14}$$

This non-equilibrium part of the actual flux can be written in the form

$$\mathbf{G}_v(x) = -\nu \frac{d\mathbf{F}}{dx} \quad \text{with} \quad \nu = \frac{\Delta x}{2} \left[ (\mathbf{J}_{\mathbf{G}^+\mathbf{m}} + \mathbf{J}_{\mathbf{G}^-\mathbf{m}}) [\mathbf{J}_{\mathbf{Fm}}]^{-1} \right]. \tag{15}$$

The *constitutive* relation above has a full 3x3 matrix of viscosity coefficients relating the flux of mass, momentum, and energy to the gradient of mass, momentum, and energy, but the important thing to note is the dependence of all the viscosity coefficients on the cell size in the first order scheme analyzed. Macrossan [28], develops this argument for the special case of the flux of momentum depending on the gradient of momentum in the context of a first order EFM applied to a perfect gas and obtains a coefficient of viscosity which is again seen to be linearly dependent on the cell size. Thus the near-equilibrium flow technique is expected to simulate equilibrium flow in the limit of the coefficient of viscosity, as, *e.g.*, in (15), going to zero. This can be achieved either by letting the cell size go to zero (expensive) or by making the method increasingly higher order accurate in space, *i.e.*, make the coefficient of viscosity of the method depend only on higher powers of the cell size. The latter approach is pursued in the next section.
25 May 1993

## 3.3 Second Order Schemes

An extension of the above first order scheme to a higher order accuracy requires care. While a second order scheme can be easily derived by averaging the individual flux terms in (7) symmetrically over adjacent neighbors, such a scheme does not use the flux-splitting nature of (7), and can be shown to be unstable. (Steep gradients cause a pathological oscillatory behavior of the numerical solution.) A second order scheme can also be obtained by calculating the flux terms in (7) using the backward slope at $x$ and the forward slope at $x + \Delta x$. This scheme is again seen to develop spurious oscillations when there are steep gradients. We note here that extensive computations with the first order scheme reiterate its physical basis: no spurious oscillatory behavior occurs in handling steep gradients (see Sec. 3.2). In light of this, the instability of the above second order scheme can be explained as follows: the second order accuracy was achieved by linear interpolation of the primary variables between the centroidal values. In so doing, the interpolated values at the cell boundaries — from the left and the right — develop spurious variations. They are spurious in that they are not in conformity with the variations as given by the kinetic (first order) process. These non-physical variations grow, particularly so in regions of steep gradients, resulting in the instability. A successful means for constructing a second order scheme, then, is to linearly interpolate the variables in such a manner as to preserve the first order monotonicities: this precludes the generation of the spurious variations at the cell boundaries. As discussed in the review article by Yee [19], introduced perhaps by van Leer [20], this can be achieved by using some kind of a limiting procedure: imposing constraints on the gradients of the primary dependent variables gives rise to *slope limiter* schemes. Imposing constraints on the gradients of the fluxes themselves gives rise to *flux limiters* [19]. Both, a slope limited second order scheme and a flux limited second order scheme are studied. The limiting procedure used in the two cases is the same, and is the popular *minmod* limiter.

The minmod limiter is defined as the binary operator:

$$\text{min\_mod}(p, q) = sgn(p) \begin{cases} 0 & \text{if } \text{sgn}(p) \neq \text{sgn}(q) \\ \min\{|p|, |q|\} & \text{if } \text{sgn}(p) = \text{sgn}(q) \end{cases} \qquad (16)$$

where $\text{sgn}(p)$ is the sign of $p$ and $|p|$ is the absolute value of $p$. In the present usage, $p$ and $q$ are the values of the slopes at the centroid of a cell — $p$ being the backward slope and $q$ the forward slope. The full minmod limiting procedure simply consists of applying the above binary operation to each of the cells at any given time step to obtain the slopes for any relevant quantity at each of the cell centroids. In both the flux limiter and slope limiter schemes to be presented, the aim is to obtain second order accurate approximations for the four split-fluxes in (7). Issues of accuracies of time integration will be discussed in the next section.

First the flux limiter scheme: the split-fluxes are calculated at the centroids and then are linearly interpolated using the minmod limiter to estimate their values at the cell boundaries as follows.

$$\begin{aligned}\mathbf{G}^+(x + \frac{\Delta x}{2}, t) =& \mathbf{G}^+(x, t) + \frac{1}{2}\text{min\_mod}\left(\Delta_{bck}\mathbf{G}^+(x, t), \Delta_{fwd}\mathbf{G}^+(x, t)\right) \\ \mathbf{G}^-(x + \frac{\Delta x}{2}, t) =& \mathbf{G}^-(x + \Delta x, t)- \\ & \frac{1}{2}\text{min\_mod}\left(\Delta_{bck}\mathbf{G}^-(x + \Delta x, t), \Delta_{fwd}\mathbf{G}^+(x + \Delta x, t)\right)\end{aligned} \quad (17)$$

where $\mathbf{G}^+(x, t)$ has been used for $\mathbf{G}^+(\mathbf{n}^+(x, t))$, *etc.*, all consistent with the kinetic flux-splitting idea and $\Delta_{fwd}y(x, t) = y(x + \Delta x, t) - y(x, t)$, *etc.*

In the slope limiter strategy, the primary dependent variables **m** at the centroids are linearly interpolated using the minmod limiter to approximate their values at the boundaries, and the split-fluxes are calculated using these values of the primary variables.

$$\mathbf{m}(x + \frac{\Delta x}{2}, t) = \mathbf{m}(x, t) + \frac{1}{2}\text{min\_mod}\left(\Delta_{bck}\mathbf{m}(x, t), \Delta_{fwd}\mathbf{m}(x, t)\right)$$

$$\mathbf{G}^+(x + \frac{\Delta x}{2}, t) = \mathbf{G}^+(\mathbf{n}^+(x + \frac{\Delta x}{2}, t)) \quad (18)$$

$$\mathbf{G}^-(x + \frac{\Delta x}{2}, t) = \mathbf{G}^-(\mathbf{n}^-(x + \frac{\Delta x}{2}, t))$$

where the notation is as in (17).

### 3.4 Time Integration

If the above spatially second order schemes are to be used in an unsteady problem, it is important to make sure that the time integration is at least second order accurate, since otherwise, the time integration errors are likely to dominate the spatial errors. In fact, the time integration errors are directly coupled to the spatial errors, owing to the CFL stability criterion. To simplify matters, a high degree of accuracy of the time integration is ensured by using fourth order Runge-Kutta with a time step well below the CFL stability limit. This is done for both the first order and second order spatially accurate schemes. Since the present interest lies in applying these schemes to unsteady problems, the high degree of accuracy of time integration is appropriate: the computations can be continued *much* longer before the accumulated time errors become significant.

## 4. The Nine-Velocity Gas

Since, the test problem and the example application in the next two sections use the nine-velocity gas in 2-D, we digress here momentarily to briefly describe the model. The model [21,22,23] consists of a large number of identical hard-sphere (disk) particles, each of which take on one of only nine allowed velocities. But for that simplification of discretizing the velocity space, the nine-velocity gas is much like a monatomic gas with a hard-sphere potential, and consists of free-flight interrupted by instantaneous collisions with other particles. The collisions, in addition to conserving mass, momentum, and energy individually, are closed under the velocity set, *i.e.*, collisions always result in post-collision velocities which are one of the nine-allowed. Fig. 1 shows the allowed velocities in the model and the four different types of collisions possible in the model, all of them preserving mass, momentum, and energy. Collision type 3 is unique in that the pre-collision speeds are different from the post-collision speeds, and this provides the crucial mechanism for equilibration between the various particle speeds.

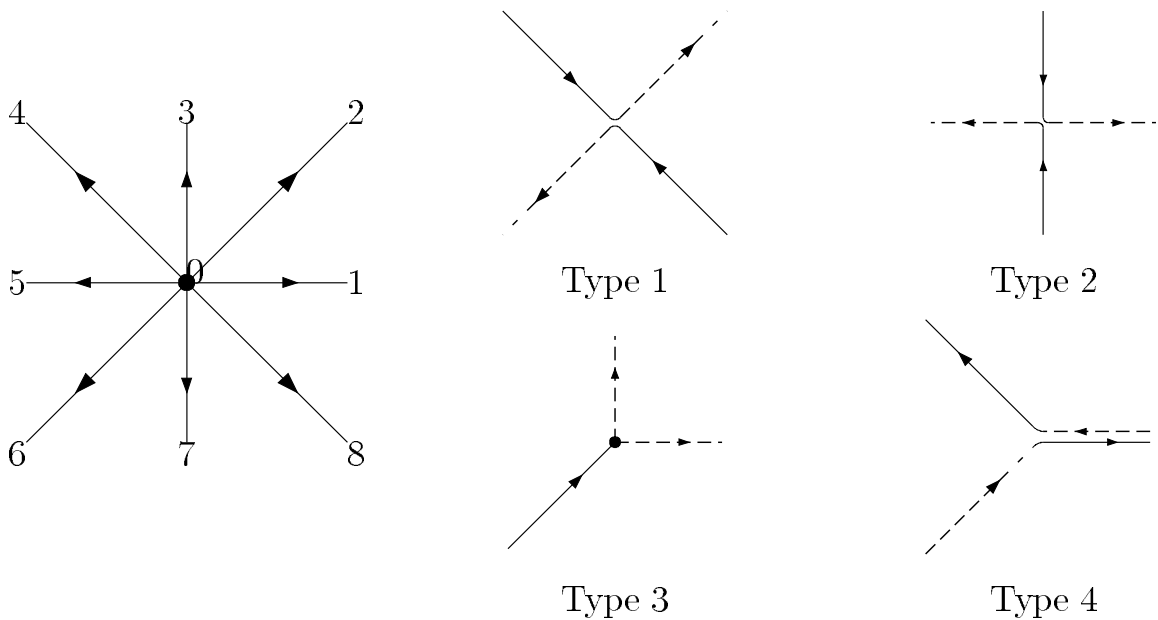

FIG. 1 The nine velocities allowed in the model comprising three different speeds and the four different types of binary collisions, all of which conserve mass, momentum, and energy, possible between identical hard-sphere particles taking on the allowable velocities.

The equilibrium relations (1), representing detailed balancing of the collisions in the model, are

$$n_0 n_2 = n_1 n_3,$$
$$n_0 n_4 = n_3 n_5,$$
$$n_0 n_6 = n_5 n_7,$$
$$n_0 n_8 = n_7 n_1,$$
$$n_1 n_5 = n_3 n_7.$$

The vector $\mathbf{F}$ of mass, momentum and energy is

$$F_1 = n = n_0 + n_1 + n_2 + n_3 + n_4 + n_5 + n_6 + n_7 + n_8,$$
$$F_2 = nu = n_1 + n_2 - n_4 - n_5 - n_6 + n_8,$$
$$F_3 = nv = n_2 + n_3 + n_4 - n_6 - n_7 - n_8,$$
$$F_4 = ne_t = n_1 + n_3 + n_5 + n_7 + 2(n_2 + n_4 + n_6 + n_8).$$

The Jacobian $\mathbf{J_{Fm}}$ of $\mathbf{F}$ with respect to $\mathbf{m}$, a vector of four of the $n$s can be obtained from the above relations (omitted here for reasons of bulk). The split fluxes for the $x$-direction are

$$G_1^+ = n_1 + n_2 + n_8,$$
$$G_2^+ = n_1 + n_2 + n_8,$$
$$G_3^+ = n_2 - n_8,$$
$$G_4^+ = n_1 + 2 * (n_2 + n_8),$$
$$G_1^- = n_4 + n_5 + n_6,$$
$$G_2^- = -n_4 - n_5 - n_6,$$
$$G_3^- = n_4 - n_6,$$
$$G_4^- = n_5 + 2(n_4 + n_6).$$

## 5. A Comparison of The First and Second Order Schemes

The accuracies of the first order and the two second order schemes as obtained in computations are compared by running a test problem with the nine-velocity gas. The time at which the methods are compared are such that the errors due to the time integration are negligible. The integral over the spatial domain of mass, momentum, and total energy is conserved (in time) to better than one part in a million in all the schemes. The order of accuracy of the method is estimated by looking at a global error measure $E$ given by

$$E(t) = \int_{Domain} y(x, t; \Delta x) - \int_{Domain} y(x, t; \Delta x = 0). \tag{19}$$

$y(x, t; \Delta x)$ is the solution obtained by a numerical scheme with a step size $\Delta x$ for a problem, the exact solution for which is given by $y(x, t; \Delta x{=}0)$. Since the exact solution $y(x, t; \Delta x{=}0)$ is not known analytically, *Richardson's deferred approach to the limit* [24] is used to estimate the exact value: a rational function extrapolation is used on a sequence of solutions with decreasing $\Delta x$ to estimate the solution at $\Delta x{=}0$. The method is $n^{th}$ order accurate if $E = O((\Delta x)^n)$.

25 May 1993

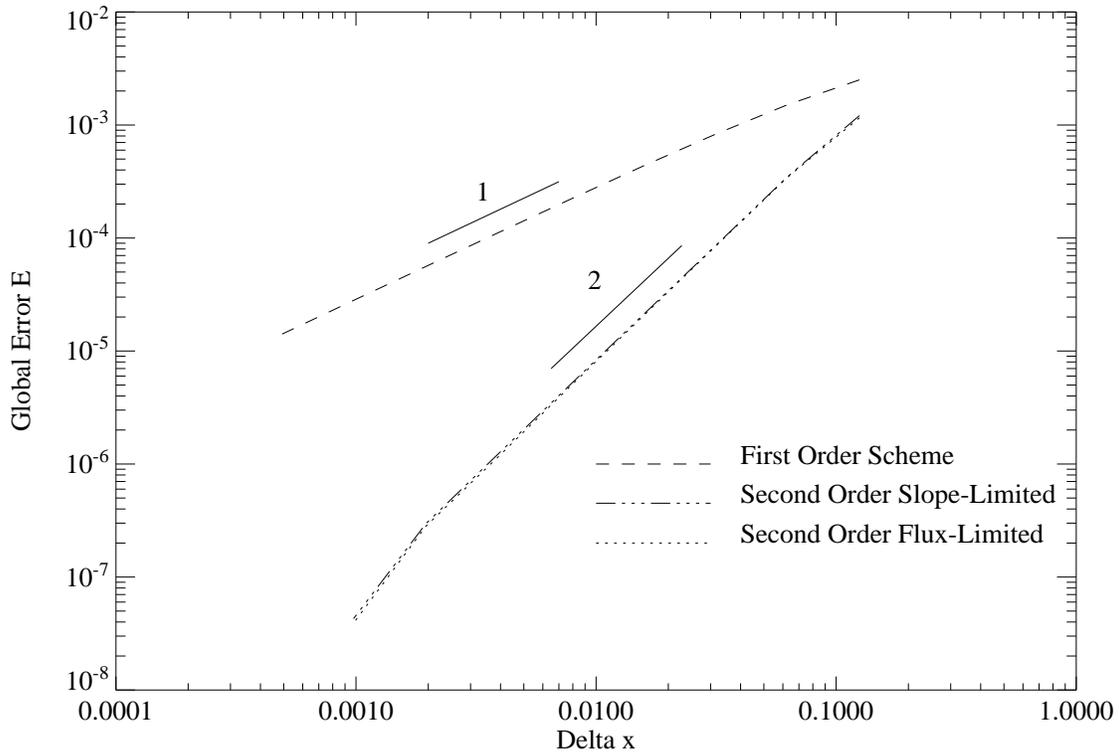

FIG. 2 Variation of the global error in kinetic energy with spatial discretization, at a particular time for the first order scheme and the two second order schemes. Slopes +1 and +2 are shown in solid line for comparison.

The test problem is set up on a periodic domain (0,1) so that the effects of boundaries are eliminated. The initial condition on the periodic domain is a sinusoidal velocity distribution superimposed over a uniform state: $\rho(x,0) = \rho_0$, $e(x,0) = e_0$, $u(x,0) = u_0 + u_1 \sin(2\pi x)$. The time $t$ at which the error analysis is done is chosen to be smaller than the time of formation of discontinuities, which is determined by the initial conditions. The quantity $y$ in (19) used in the error analysis is the non-dimensional kinetic energy $\frac{1}{2}\rho u^2$. (Density, $\rho$, is non-dimensionalized by the average density over the domain and the velocity, $u$, by the unit speed, $q$ in the model.) Fig. 3 shows the scaling of the global errors with the spatial discretization for the three different schemes discussed previously. The first order accuracy of the basic scheme and the second order accuracy of both the flux limited and slope limited schemes are clearly verified. The close correspondence between the flux limiter and slope limiter schemes, shows that neither of them has a clear advantage over the other.

## 6. An Example Application

As an example, the flux-limited second order method is applied to a shock tube problem, using the nine-velocity gas described in Sec. 4. The initial condition corresponds to a high density, high pressure driver section and a low density, low pressure driven section, separated by a diaphragm which ruptures at $t = 0$. Initially, the density ratio is 5, and the specific energy ratio 2.5. The gas in the two regions are the same, the nine-velocity gas, and is at rest. The end walls are modeled by mirror image sites across the actual wall location. This results in specular walls, which therefore act as no-flow (zero mass flux), adiabatic (zero total energy flux) boundaries. 256 cells were used to simulate the shock tube, and the three primary variables, **m** in Sec. 2 & 3, were the population densities $n_0$, $n_1$, and $n_3$. The time evolution of density is shown using a grey-level coding on the $x - t$ plane in Fig. 4, and the density, specific energy, and $u$-velocity at a particular time in Fig. 5. The slight variation of $u$-velocity across the contact surface reflects the kinematic dependence of the thermodynamics of discrete-velocity gases: since the pressure depends on the flow speed as well, the jump across the contact surface, which is such that the pressure across it is constant, results in a change of velocity. The interaction of the various types of waves — the shock wave, the rarefaction fan, and the contact surface — can be seen in Fig. 4. The shock speeds in that figure concur with that obtained from the model Euler equations, (6), with the jumps in Fig. 5 satisfying the jump conditions (of (6)).

## 7. The Equilibrium Flow Technique Compared to Other Lattice Gas Methods

The features of lattice gases which make them interesting and popular are preserved in the near-equilibrium flow schemes introduced. First, the interactions are local, and thus these methods are as parallelizable as other lattice gas methods. Next, subject only to the CFL criterion, these schemes are robust with little unphysical behavior.

As compared to the full lattice gas simulations, the relative merits of the near-equilibrium flow technique are

1. Since it is not a stochastic process, simulations using this technique are relatively noise-free.

2. The need for an exclusion principle to ease computation is obviated. Thus the classically unphysical effects of an exclusion principle are eliminated.

3. Schemes to enhance collisions — in some instances, they violate basic conservations, leading to a loss of *universality of equilibrium distributions!* [27] — are also obviated since the technique models the *infinite collision rate* limit.

4. Spurious conservations, as of staggered momenta [14], are not present in this mode of computation.

5. It is speculated that simulations of flows with a much higher Reynolds number are possible with this method, given the same computer resources. The quantitative interpretation of the flows will perhaps be complicated by the nature of the viscosity coefficient (*e.g.*, (15)).

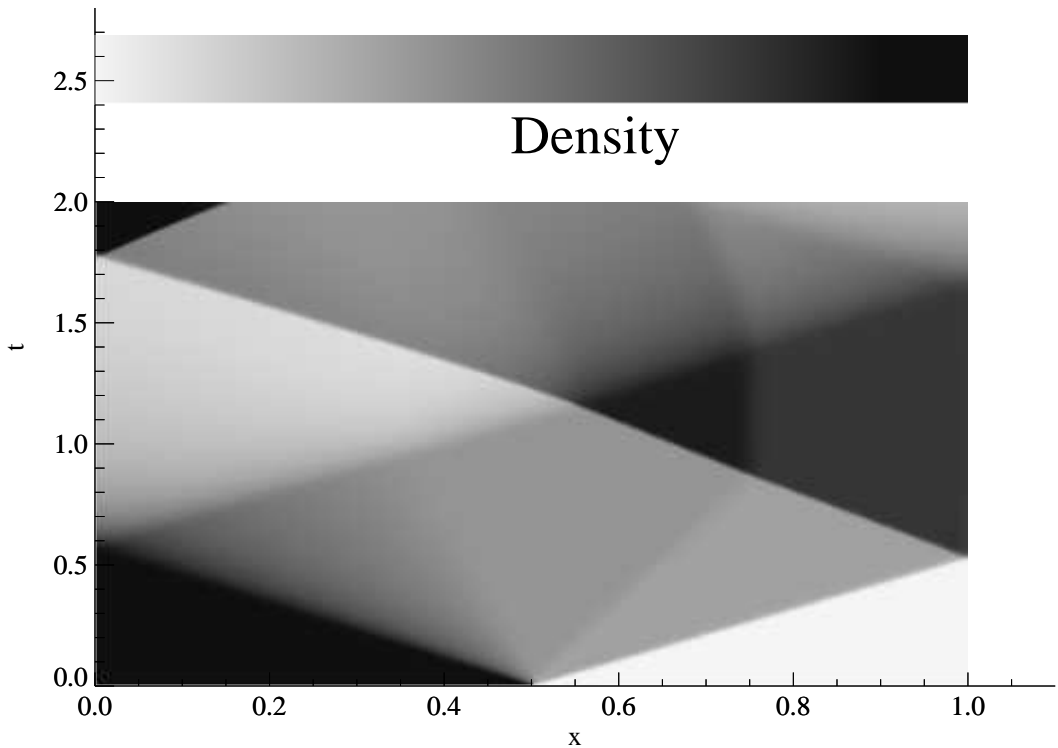

FIG. 5 Density variations indicated on the $x-t$ plane for the shock tube problem. The interactions of the shock wave, the rarefaction wave, the contact surface, and the end-walls are all captured by the method. The computation here was carried out using a nine-velocity gas.

6. If a larger number of velocites are to be included in the model, while the LGA schemes are likely to run into problems of too big a neighborhood, and/or too large a look-up table, the present method would not be affected much, but for an increased computation in each cell.

7. Extension to computations in 2 and 3 dimensions poses no new problems.

The disadvantages of the near-equilibrium flow technique compared to LGA are,

1. Aspects of long-time velocity auto-correlations and many-body correlations cannot be studied, since they are all thrown away by the Boltzmann approximation.

2. By using floating point numbers, it lacks a very attractive feature of CA. This may preclude implementations of this technique on the special CA computers being developed [29].

3. One has to speak of the accuracy of the method unlike in the CA universe.

To put it in the context of the Lattice Boltzmann Equation (LBE) approach, the near-equilibrium flow technique is the *infinite collision rate* limit of the *full* LBE. Computing collisions which constitute the right hand side of the LBE is a computationally intensive task [26]. The present method does not deal with collisions, since an infinite collision rate is *implicit* in the formulation. The prevalent schemes for LBE, however, use a simplified model for the collisions, a relaxation to equilibrium, like the Bhatnagar-Gross-Krook (BGK) collision model for the *actual* Boltzmann equations [30], but with other restrictions, *e.g.*, (1) since the exact equilibrium distributions are not known, they are approximated, (2) since there is no more a physical basis, the form of relaxation is arbitrary, *etc.*. A price to be paid for such simplifications, however, is not knowing when the simplified models of collision becomes insufficient or unreasonable. This is evidenced even in the studies of the BGK approximation of the actual Boltzmann equations [25]. Since the present near-equilibrium flow techniques are based on the idea of local thermodynamic equilibrium, no such problems arise in their usage.

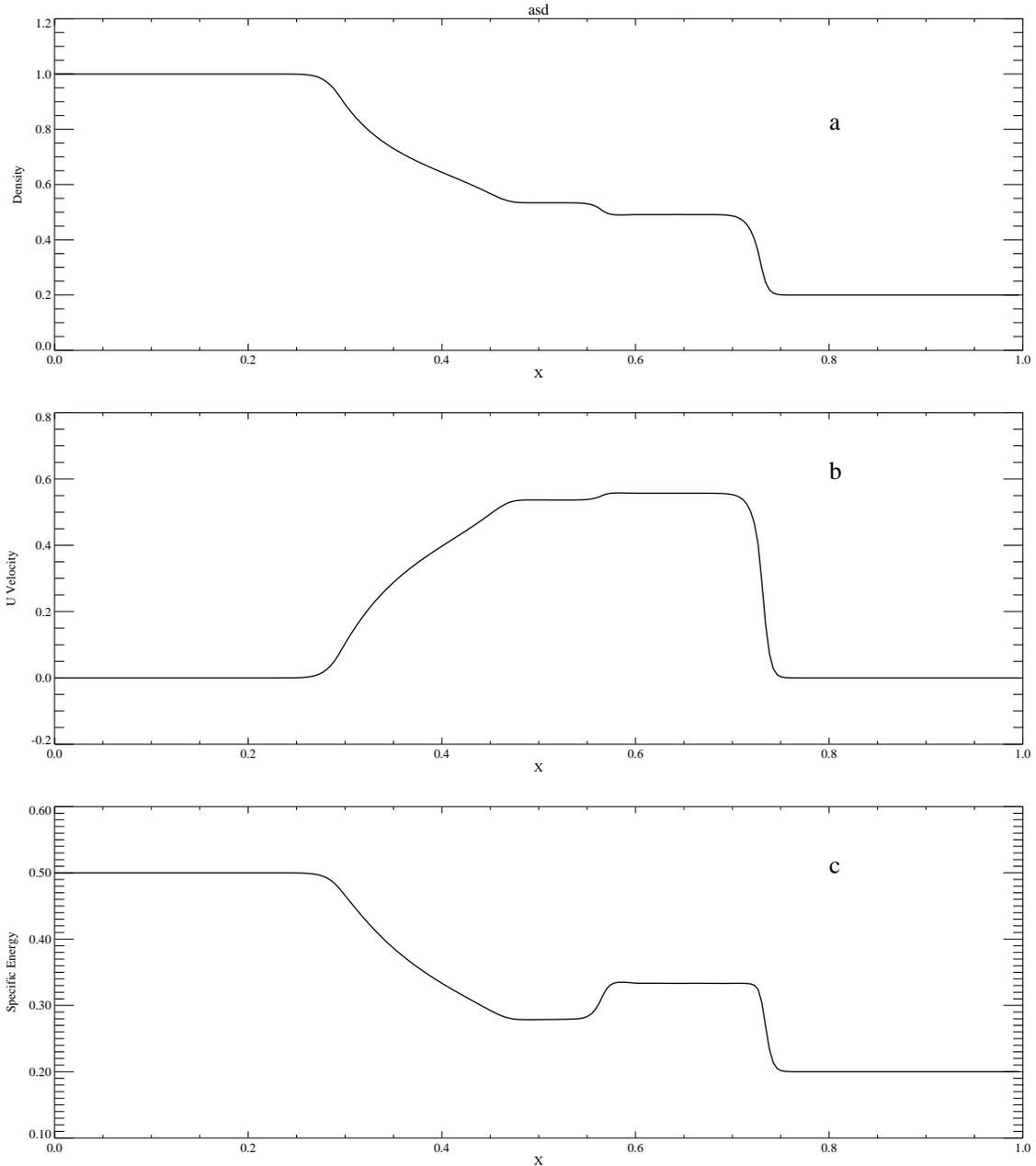

FIG. 6 The density, *u*-velocity, and specific-energy variations in the shock tube at t=0.5. These plots are from the same shock tube computation, the $x - t$ variation of density for which was indicated in Fig. 3.

## 8. Conclusion

Using ideas from the equilibrium flux method — a kinetic-based finite volume method for compressible ideal gas flows — in the context of discrete-velocity gases, we were able to develop a near-equilibrium flow technique for discrete-velocity gases. With a simple transformation of variables in some of the equations, the method carries through for lattice gases. While these techniques retain the attractive features of lattice gases — simplicity and parallel evolution, they represent a large improvement over the currently used LGA and LBE techniques, in that the evolution is based on the inviscid, non-heat conducting limit.

## Acknowledgements

One of the authors (BTN) would like to thank Dr. B. Sturtevant and Dr. J. E. Broadwell for their support and helpful discussions in carrying out the work. This work was supported in part by the NSF under Cooperative Agreement No. CCR-8809615 and by the AFOSR under grant AFOSR-89-0369.

25 May 1993

– 17 –

# Appendix

The detailed relations for the nine-velocity model (see Fig. 1) are given here, as an example. The equilibrium relations (1) are

$$n_0 n_2 = n_1 n_3,$$
$$n_0 n_4 = n_3 n_5,$$
$$n_0 n_6 = n_5 n_7,$$
$$n_0 n_8 = n_7 n_1,$$
$$n_1 n_5 = n_3 n_7.$$

The vector $\mathbf{F}$ of mass, momentum and energy is

$$F_1 = n = n_0 + n_1 + n_2 + n_3 + n_4 + n_5 + n_6 + n_7 + n_8,$$
$$F_2 = nu = n_1 + n_2 - n_4 - n_5 - n_6 + n_8,$$
$$F_3 = nv = n_2 + n_3 + n_4 - n_6 - n_7 - n_8,$$
$$F_4 = ne_t = n_1 + n_3 + n_5 + n_7 + 2(n_2 + n_4 + n_6 + n_8).$$

The Jacobian $\mathbf{J_{Fm}}$ of $\mathbf{F}$ with respect to $\mathbf{m}$, a vector of four of the $n$s can be obtained from the above relations (omitted here for reasons of bulk). The split fluxes for the $x$-direction are

$$G_1^+ = n_1 + n_2 + n_8,$$
$$G_2^+ = n_1 + n_2 + n_8,$$
$$G_3^+ = n_2 - n_8,$$
$$G_4^+ = n_1 + 2 * (n_2 + n_8),$$
$$G_1^- = n_4 + n_5 + n_6,$$
$$G_2^- = -n_4 - n_5 - n_6,$$
$$G_3^- = n_4 - n_6,$$
$$G_4^- = n_5 + 2(n_4 + n_6).$$

25 May 1993